\documentclass[10pt]{article}

\usepackage{graphicx}
\usepackage{dcolumn}
\usepackage{bm}

\begin{document}
\begin{center}
{{\LARGE \bf Hot and Dense Hadron Gas (HG): A New Excluded-volume approach }} \\
\bigskip
{\large \bf S. K. Tiwari \footnote{email: sktiwari4bhu@gmail.com}}, and {\large \bf C. P. Singh}\\
{Department of Physics, Banaras Hindu University, Varanasi-05, India} 
\bigskip
\end{center}

\begin{abstract}
We formulate a thermodynamically consistent equation of state (EOS), based on excluded-volume approach, for a hot, dense hadron gas (HG). We calculate various thermodynamical quantities of HG and various hadron ratios and compare our model results with the results of other excluded-volume models and experimental data. We also calculate various transport coefficients such as $\eta/s$ etc. and compare them with other HG model results. Furthermore, we test the validity of our model in getting the rapidity spectra of various hadrons and the effect of flow on them is investigated by matching our predictions with the experimental data.  
\end{abstract}


\section{Introduction}
The search for any unambiguous signal of quark-gluon plasma (QGP) in heavy-ion collisions is motivated by properly investigating the behaviour and properties of hot and dense HG \cite{Singh:1993}. Here we formulate a new thermodynamically consistent excluded-volume model where we assign a finite hard-core volume to each baryon but mesons in the theory can easily overlap, fuse and interpenetrate into each other. Earlier, we have used this model successfully in obtaining the conjectured QCD phase boundary and thus determining precisely the location of QCD critical end point \cite{Singh:2009,Srivastava:2010}. We calculate various thermodynamical quantities like number density etc. of HG and compare our model results with that of URASiMA event generator \cite{Sasaki:2001}. We use our freeze-out picture \cite{Tiwari:2012} for calculating various hadron ratios and compare our results with the experimental data and various excluded-volume models. In order to make the discussion complete, we further derive $\eta/s$ etc. from our model and compare them with other models. Further, we extend our model to deduce the rapidity as well as transverse mass spectra of hadrons and compare them with the experimental data available in order to illustrate the role of flow present in the fluid. Finally, we give summary of this work.

\section{Model Descriptions}

Recently, we have proposed a thermodynamically consistent excluded-volume model for a hot, dense HG \cite{Singh:2009,Srivastava:2010,Tiwari:2012}. The attractive interaction between baryons and mesons is realized by including the baryon and meson resonances in our model calculation. The repulsive interaction between baryons is modelled via giving an equal and finite size to each baryon. Mesons can fuse and interpenetrate into each other so, they are treated as pointlike particles. Using quantum statistics, the grand canonical partition function for baryons can be written as follows :   

\begin{equation}
ln Z_i^{ex} = \frac{g_i}{6 \pi^2 T}\int_{V_i^0}^{V-\sum_{j} N_j V_j^0} dV
\int_0^\infty \frac{k^4 dk}{\sqrt{k^2+m_i^2}} \frac1{[exp\left(\frac{E_i - \mu_i}{T}\right)+1]}
\end{equation}
where $g_i$ is the degeneracy factor of ith species of baryons,$E_{i}$ is the energy of the particle ($E_{i}=\sqrt{k^2+m_i^2}$), $V_i^0$ is the eigen volume assigned to each baryon of ith species. Apparently, our approach is more simple in comparison to other thermodynamically consistent excluded-volume approach \cite{Rischke:1991} which often involve transcendental expressions and are difficult to solve. We determine chemical freeze-out temperature and chemical potential by fitting our results with some experimental data and then use them in determining all other quantities and ratios. 


 The rapidity distributions of baryons using thermal source can be written as follows \cite{Tiwari1:2012} :

\begin{equation}
\Big(\frac{dN}{dy}\Big)_{th}=\frac{g_iV\lambda_i}{2\pi^2}\;\Big[(1-R)-\lambda_i\frac{\partial{R}}{\partial{\lambda_i}}\Big]
exp\left(\frac{-m_i\;coshy}{T}\right)\Big[m_i^2T+\frac{2m_iT^2}{coshy}+\frac{2T^3}{cosh^2y}\Big].
\end{equation}

where $m_i$ is the mass of the ith species. $V$ is the freeze-out volume of the system. The rapidity spectra of hadrons with the effect of flow can be calculated by using following formula \cite{Tiwari1:2012} :

\begin{eqnarray}
\frac{dN_i}{dy}=\int_{-\eta_{max}}^{\eta_{max}} \Big(\frac{dN_i}{dy}\Big)_{th}(y-\eta)\;d\eta,
\end{eqnarray}
where $\eta_{max}$ is a free parameter related with the longitudinal flow velocity ($\beta_L$) \cite{Tiwari1:2012}.

Our excluded-volume approach involves hard-core repulsion arising between two baryons but mesons do not possess any such repulsion. We have taken an equal volume $\displaystyle V^{0}=4\pi r^{3}/3$ for each baryon with a hard-core radius $r=0.8\; fm$. We have also taken all baryons, mesons and their resonances having masses upto $2\;GeV/c^{2}$ in our calculation of HG pressure. We have also imposed the condition of strangeness neutrality by considering $\sum_{i}S_{i}(n_{i}^{s}-\bar{n}_{i}^{s})=0$ where $S_{i}$ is the strangeness of ith hadron.

\section{Results and Discussions}

\begin{figure}
\begin{center}
\includegraphics[scale=0.3]{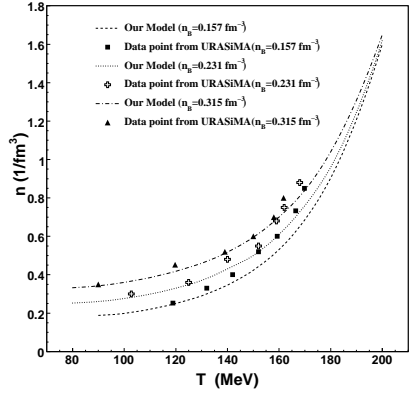}
\caption{Variation of total number density of hadrons with respect to temperature at constant net baryon density. lines show our model calculation and points are the data calculated by Sasaki using URASiMA event generator.}
\label{fig9}
\end{center}
\end{figure}

\begin{figure}
\begin{center}
\includegraphics[scale=0.3]{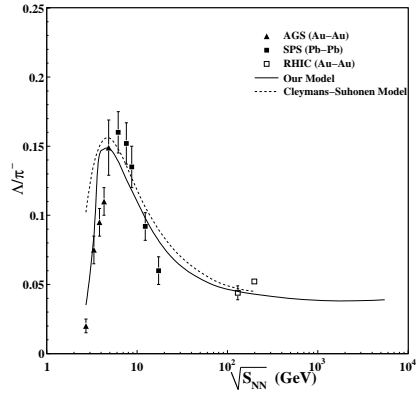}
\caption{The energy dependence of $\Lambda/\pi^{-}$ ratio. Dashed line is the result of Cleymans-Suhonen model \cite{Cleymans:1987}. Symbols are the experimental data \cite{Tiwari:2012}. RHIC data are at mid-rapidity.}
\label{fig9}
\end{center}
\end{figure}

In Fig. 1, we have plotted the variation of number density of hadrons with respect to temperature at fixed net baryon density. Our model results show a very good agreement with the results of Sasaki \cite{Sasaki:2001} except at higher $T$. Figure 2 represents the variation of $\Lambda/\pi^{-}$ ratio with $\sqrt{s_{NN}}$. We find that our model calculation gives much better fit to the experimental data at all energies in comparison to Cleymans-Suhonen model \cite{Cleymans:1987}. Figure 3 depicts the variation of $\eta/s$ with respect to temperature as obtained in our model for HG \cite{Tiwari:2012} having a baryonic hard-core size $r=0.5$ fm, and compared the results with those of Gorenstein $et\; al.$ \cite{Gorenstein:2008}. We find that near the expected QCD phase transition temperature ($T_{c}=170-180$ MeV), $\eta/s$ shows a lower value in our HG model than the value in other model. In Fig. 4, we show the rapidity distributions of $\pi^+$ for central $Au+Au$ collisions at $\sqrt{s_{NN}}=200\; GeV$. Dotted line shows the distribution of $\pi^+$ due to stationary thermal source. Solid line shows the rapidity distributions of $\pi^+$ after the incorporation of longitudinal flow in our thermal model and results give a good agreement with the experimental data \cite{Tiwari1:2012}.

\begin{figure}
\begin{center}
\includegraphics[scale=0.3]{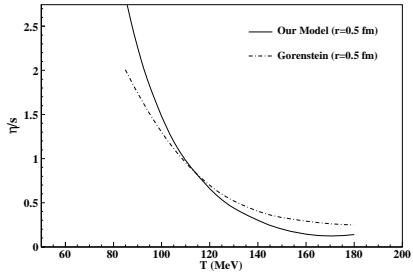}
\caption{Variation of $\eta/s$ with temperature for $\mu_{B}=0$ in our model and a comparison with the results obtained by Gorenstein $et\; al.$ \cite{Gorenstein:2008}.}
\label{fig9}
\end{center}
\end{figure}

\begin{figure}
\begin{center}
\includegraphics[scale=0.3]{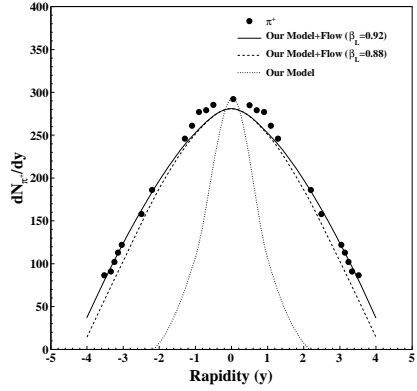}
\caption{Rapidity distributions of $\pi^+$ at $\sqrt{s_{NN}}= 200 GeV $. Dotted line shows the rapidity distribution calculated in our thermal model. Solid line and dashed line show the results obtained after incorporating longitudinal flow in our thermal model. Symbols are the experimental data \cite{Tiwari1:2012}.}
\label{fig9}
\end{center}
\end{figure}

\section{Summary}
In summary, we find that our model results mostly show very close agreement with those of Sasaki, although the two approaches are completely different in nature. Our model calculations for the particle ratios describe the experimental data very well. Transport quantities are also successfully described by our model. The rapidity distributions which essentially are dependent on thermal parameters, also show a systematic behaviour and their interpretations most clearly involve the presence of a collective flow involved in the final description of the fireball. 

\section{Acknowledgment}
SKT is grateful to Council of Scientific and Industrial Research (CSIR), New Delhi for providing a research grant.


\noindent
\begin{thebibliography}{50}
\medskip

\bibitem{Singh:1993}
 C.~P.~Singh, Phys. Rep. {\bf 236}, 147 (1993); Int. J. Mod. Phys. A {\bf 7}, 7185 (1992).




\bibitem{Singh:2009}
 C. P. Singh, P. K. Srivastava and S. K. Tiwari, Phys. Rev. D {\bf 80}, 114508 (2009); Phys. Rev. D {\bf 83}, 039904(E) (2011).




\bibitem{Srivastava:2010}
P. K. Srivastava, S. K. Tiwari, and C. P. Singh, Phys. Rev. D {\bf 82}, 014023 (2010); Nucl. Phys. A {\bf 862-863 C}, 424-426 (2011).



\bibitem{Sasaki:2001}
N. Sasaki, Progress of theoretical Physics {\bf 106}, 783 (2001).




\bibitem{Tiwari:2012}
S. K. Tiwari, P. K. Srivastava, and C. P. Singh, Phys. Rev. C {\bf 85}, 014908 (2012). 



\bibitem{Rischke:1991}
D. H. Rischke, M. I. Gorenstein, H. St\"{o}cker and W. Greiner, Z. Phys. C {\bf 51}, 485 (1991).





\bibitem{Tiwari1:2012}
S. K. Tiwari, P. K. Srivastava, and C. P. Singh, arXiv:1202.4852[hep-ph].





\bibitem{Cleymans:1987}
J. Cleymans and E. Suhonen, Z. Phys. C {\bf 37}, 51 (1987).




\bibitem{Gorenstein:2008}
M. I. Gorenstein, M. Hauer, and O. N. Moroz, Phys. Rev. C {\bf 77}, 024911 (2008).














\end{thebibliography}
\end{document}